\documentclass[11pt]{article}
\usepackage{graphicx}

\newcommand{\BABARPubYear}    {06}

\newcommand{\BABARConfNumber} {33}
\newcommand{\SLACPubNumber} {11983}
\RequirePackage{xspace}
\usepackage{relsize}
\mathchardef\Upsilon="7107
\def\Y#1S{\ensuremath{\Upsilon{(#1S)}}\xspace}
\def\FourS {\Y4S}
\def\pep2{PEP-II}
\def\Dz      {\ensuremath{D^0}\xspace}
\def\Dzb     {\ensuremath{\Dbar^0}\xspace}
\def\BR         {{\ensuremath{\cal B}\xspace}}
\def\Dbar    {\kern 0.2em\overline{\kern -0.2em D}{}\xspace}
\newcommand{\gev}{\ensuremath{\mathrm{\,Ge\kern -0.1em V}}\xspace}
\newcommand{\gevcc}{\ensuremath{{\mathrm{\,Ge\kern -0.1em V\!/}c^2}}\xspace}
\newcommand{\gevccc}{\ensuremath{{\mathrm{\,Ge\kern -0.1em V^2/}c^4}}\xspace}
\newcommand{\mevcc}{\ensuremath{{\mathrm{\,Me\kern -0.1em V\!/}c^2}}\xspace}
\newcommand{\gevc}{\ensuremath{{\mathrm{\,Ge\kern -0.1em V\!/}c}}\xspace}
\def\babar{\mbox{\slshape B\kern-0.1em{\smaller A}\kern-0.1em
    B\kern-0.1em{\smaller A\kern-0.2em R}}}
 
\long\def\inst#1{\par\nobreak\kern 4pt\nobreak
    {\it #1}\par\vskip 10pt plus 3pt minus 3pt}

\begin{document}
{\pagestyle{empty}

\begin{flushright}
\babar-CONF-\BABARPubYear/\BABARConfNumber \\
SLAC-PUB-\SLACPubNumber \\
\end{flushright}
\begin{center}
\Large \bf Study of the Exclusive Initial-State Radiation Production of the $D \bar D$ System
\end{center}

\begin{center}
\small

The \babar\ Collaboration,
\bigskip

%
{B.~Aubert,}
{R.~Barate,}
{M.~Bona,}
{D.~Boutigny,}
{F.~Couderc,}
{Y.~Karyotakis,}
{J.~P.~Lees,}
{V.~Poireau,}
{V.~Tisserand,}
{A.~Zghiche}
\inst{Laboratoire de Physique des Particules, IN2P3/CNRS et Universit\'e de Savoie,
 F-74941 Annecy-Le-Vieux, France }
% ...

\end{center}\newpage

\begin{center}
\large \bf Abstract
\end{center}
A study of exclusive production of the $D \bar D$ system through initial-state radiation
is performed in a search for charmonium states, where
$D=\Dz$ or $D^+$. The $\Dz$ mesons are reconstructed
in the $\Dz \to K^- \pi^+$, $\Dz \to K^- \pi^+ \pi^0$, and
$\Dz \to K^- \pi^+ \pi^+ \pi^-$ decay modes.
The $D^+$ is reconstructed through
the $D^+ \to K^- \pi^+ \pi^+$ decay mode.
The analysis makes use of an integrated luminosity of 288.5 fb$^{-1}$ 
collected by the \babar \ experiment.
The $D \bar D$ mass spectrum shows a clear $\psi(3770)$ signal.
Further structures appear in the 3.9 and 4.1 \gevcc
regions. No evidence is found for $Y(4260)$ decays to $D \bar D$, implying an upper limit 
$\frac{\BR(Y(4260)\to D \bar D)}{\BR(Y(4260)\to J/\psi \pi^+ \pi^-)} < 7.6$ (95\% confidence level).
\vfill
\begin{center}

Submitted to the 33$^{\rm rd}$ International Conference on High-Energy Physics, ICHEP 06,\\
26 July---2 August 2006, Moscow, Russia.

\end{center}

\vspace{1.0cm}
\begin{center}
{\em Stanford Linear Accelerator Center, Stanford University, 
Stanford, CA 94309} \\ \vspace{0.1cm}\hrule\vspace{0.1cm}
Work supported in part by Department of Energy contract DE-AC03-76SF00515.
\end{center}

\newpage
} 

\setcounter{footnote}{0}
\section{Introduction}
The $Y(4260)$ structure was 
discovered in the initial-state radiation (ISR) process $e^+e^- \to \gamma_{ISR} \pi^+\pi^-J/\psi$~\cite{y4260} 
at the energy where the total $e^+e^-$
hadronic cross section shows a local minimum \cite{besR}. 
Its spin-parity assignment, $J^{PC}= 1^{--}$ is inferred because
it can be produced by a single-photon annhilation mechanism.  The measurement 
$\Gamma(Y(4260) \to e^+ e^-) \times \BR(Y(4260)\to \pi^+\pi^-J/\psi) = (5.5 \pm
1.0^{+0.8}_{-0.7}$) eV/$c^2$ 
and an upper limit estimate of the $Y(4260)$  contribution to the total cross section gives a
value for $\Gamma_{\pi \pi J/\psi}(Y(4260))$ in excess of 1 \mevcc \cite{gam_pipiPsi}, where 
  $\Gamma_{\pi \pi J/\psi}$ denotes the partial width to the $\pi^+\pi^-J/\psi$ final state.
In contrast, $\Gamma_{\pi \pi J/\psi}$ for the $\psi(3770)$ is 
($80 \pm 33 \pm 23$) keV/$c^2$~\cite{bes}, 
prompting suggestions that the $Y(4260)$ could be an exotic object~\cite{exo}. The $Y(4260)\to \pi^+\pi^-J/\psi$
decay channel indicates  $\bar c c$ content in the $Y(4260)$, which is energetically allowed to 
decay to  pairs of charmed mesons. In this paper, we report a search for 
$Y(4260) \to D \bar D$ using events accompanied by initial-state radiation at the \FourS energy.

\section{Data selection}
We use data from the \babar \ detector ~\cite{babar} at the PEP-II asymmetric-energy 
$e^+ e^-$  storage rings, located at the Stanford Linear Accelerator Center (SLAC). These 
data represent an integrated luminosity of 288.5 fb$^{-1}$ collected at  $\sqrt{s} = 10.58$ \gev, 
near the peak of the \FourS resonance, and approximately 40 MeV below
this energy.

Charged-particle momenta are measured in a tracking system consisting of a five-layer 
double-sided silicon vertex tracker (SVT) and a 40-layer central drift chamber (DCH), 
both situated in a 1.5-T axial magnetic field. An internally reflecting ring-imaging 
Cherenkov detector (DIRC) with fused-silica bar radiators provides charged-particle 
identification. Kaon and pion candidates are selected based on a likelihood calculated 
from the specific ionization in the DCH and SVT, and the Cherenkov angle measured in 
the DIRC. A CsI(Tl) electromagnetic calorimeter (EMC) is used to detect and identify 
photons and electrons. Neutral pions are reconstructed from pairs of photons, each depositing 
at least 100 MeV in the EMC and having an invariant mass between 115 and 155 \mevcc. 
Photon pairs are kinematically constrained to have the $\pi^0$ mass; pairs having a fit
probability greater than 1\% are retained as $\pi^0$ candidates. 

Candidate $e^+e^- \to  \gamma_{ISR} D \bar D$
events, where $\gamma_{ISR}$ denotes an 
ISR photon, are reconstructed using four decay channels: 
\begin{eqnarray}
 & & e^+ e^- \to \gamma_{ISR} \ \Dz \Dzb, \Dz \to K^- \pi^+, \Dzb \to K^+ \pi^- \\
 & & e^+ e^- \to \gamma_{ISR} \ \Dz \Dzb, \Dz \to K^- \pi^+ \pi^0, \Dzb \to K^+ \pi^- \\
 & & e^+ e^- \to \gamma_{ISR} \ \Dz \Dzb, \Dz \to K^- \pi^+ \pi^+\pi^-, \Dzb \to K^+ \pi^- \\
 & & e^+ e^- \to \gamma_{ISR} \ D^+ D^-, D^+ \to K^- \pi^+\pi^+, D^- \to K^+ \pi^- \pi^-.  
\end{eqnarray}
Charge conjugate decays are implicitly included. 

All charged daughter tracks of $D$ meson candidates must be well-reconstructed; however there may be one 
additional track having fewer than twelve DCH hits, having a transverse momentum 
below 0.1 \gevc, or having a distance of closest approach to the interaction point of 
greater than 1 cm in the plane transverse to the beam or 10 cm along the beam direction. 
Events having an extra $\pi^0$ candidate 
are vetoed. The tracks of each $D$ candidate are topologically constrained to come from 
a common vertex; $D$ candidates with a vertex fit probability greater than 0.1\% are retained. 
Subsequently, each $D \bar D$ pair is refit to a common vertex with a beam spot 
constraint; $D \bar D$ candidates with a fit probability of greater than 0.1\% are retained. When 
calculating kinematic properties of $D \bar D$ pairs, the energy of the $D$ mesons with 
daughter tracks that are all charged is recomputed using the nominal $D$ mass value~\cite{pdg}. 
For candidate $\Dz \to K^- \pi^+ \pi^0$ decays, a one-constraint fit to the nominal $\Dz$ mass 
is performed. Events having a $D \bar D$ pair with an invariant mass below 6.0 \gevcc 
are retained.

\begin{figure}[!htb]
\begin{center}
\includegraphics[width=9cm]{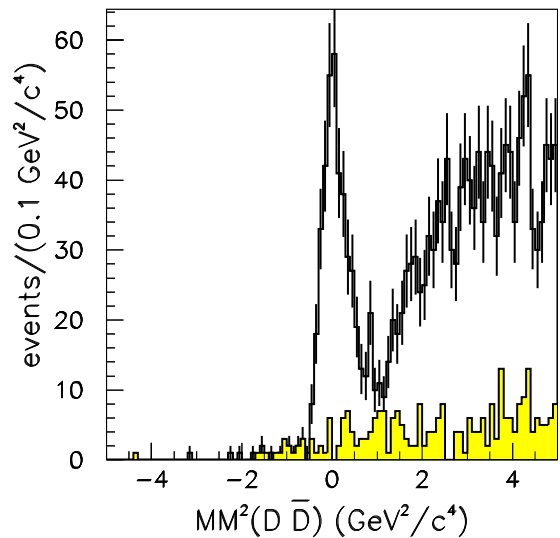}
\caption{The recoil mass squared for ISR event candidates ($MM^2$) summed over all the four
reconstructed final states. The shaded
histogram is the background distribution estimated  from $D$ and $\bar D$ mass sidebands.}
\label{fig:fig1}
\end{center}
\end{figure}

A distinguishing characteristic of $e^+e^- \to  \gamma_{ISR} D \bar D$ events is that the squared 
invariant mass of the recoil to the $D \bar D$ system ($MM^2$) is that of 
the initial-state 
photon. The peak centered on $MM^2$= 0 in Fig.~\ref{fig:fig1}, summed over all four 
reconstructed channels, provides clear evidence for exclusive 
ISR production. The shaded histogram shows a background estimate derived from the 
two-dimensional $D$ and $\bar D$ mass sidebands. 

\begin{figure}[!htb]
\begin{center}
\includegraphics[width=12cm]{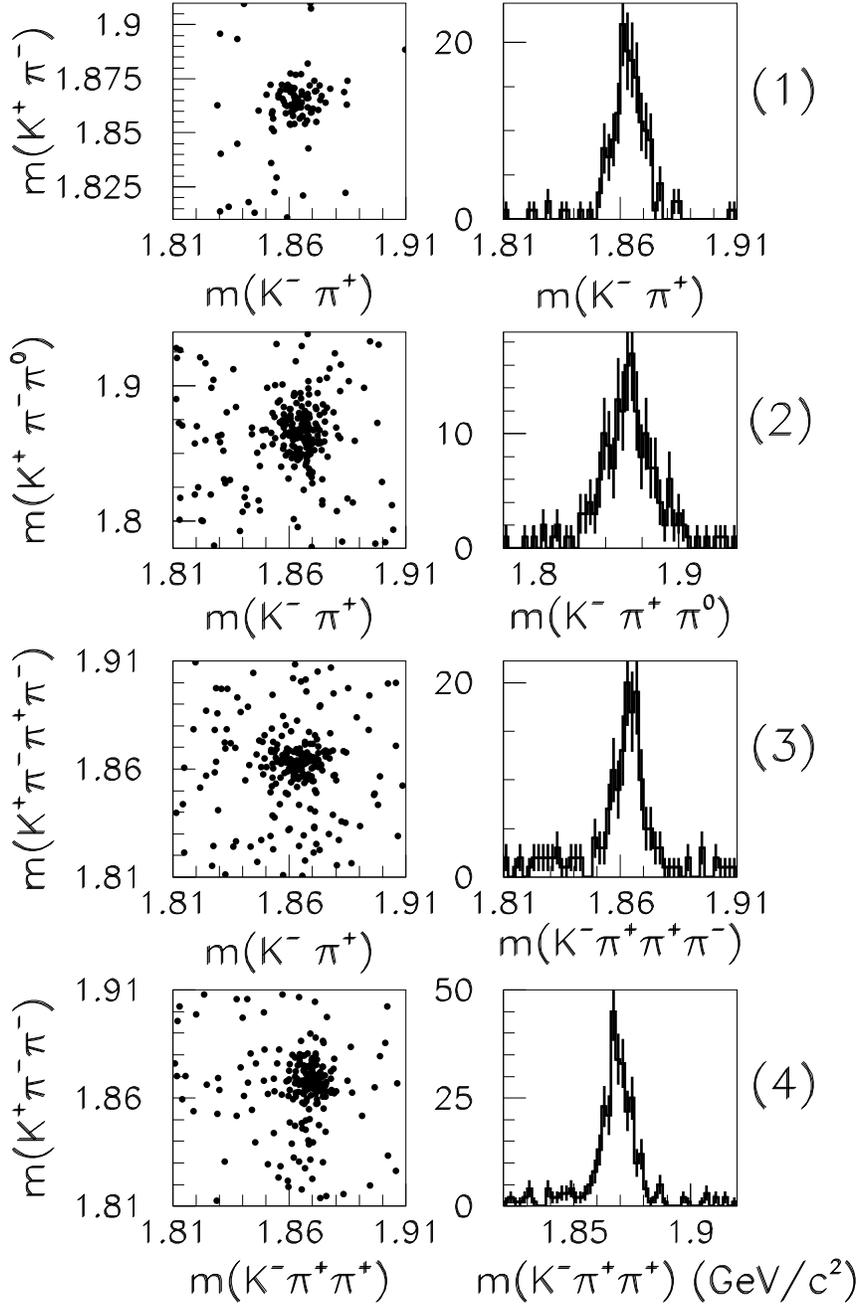}
\caption{Scatter plots for each channel and projections of the candidate $D$ mass.}
\label{fig:fig2}
\end{center}
\end{figure}

Candidate events in the ISR region, defined as 
$|MM^2| < 1.0$ \gevccc are retained. Observation of the ISR photon is not required since
these photons are emitted preferentially along the beam directions; approximately 10\% are reconstructed in the 
EMC. In Fig.~\ref{fig:fig2}, we show the mass of the
$D$ vs.~that of the $\bar D$, for each 
reconstructed channel and projections for each reconstructed $D$ decay mode.
 
The 
number of observed events for each reconstructed channel is given in
Table~\ref{tab:tab1}.
The same table also shows the resolution from a single-Gaussian fit to the $D$-candidate mass
spectra
and the experimental $D \bar D$ mass resolution at the mass of the $Y(4260)$.
The mass resolution for each channel agrees 
with values found for simulated $e^+e^- \to  \gamma_{ISR} D \bar D$ events.
A $\pm 2.5 \sigma$ signal 
region is defined for each $D$ channel, with sideband regions defined from 
--6.0 $\sigma$ to --3.5 $\sigma$ and  3.5 $\sigma$ to 6.0 $\sigma$. Background estimates for each channel 
are inferred from the 
two-dimensional $D$ and $\bar D$ sidebands.
\begin{figure}[!htb]
\begin{center}
\includegraphics[width=14cm]{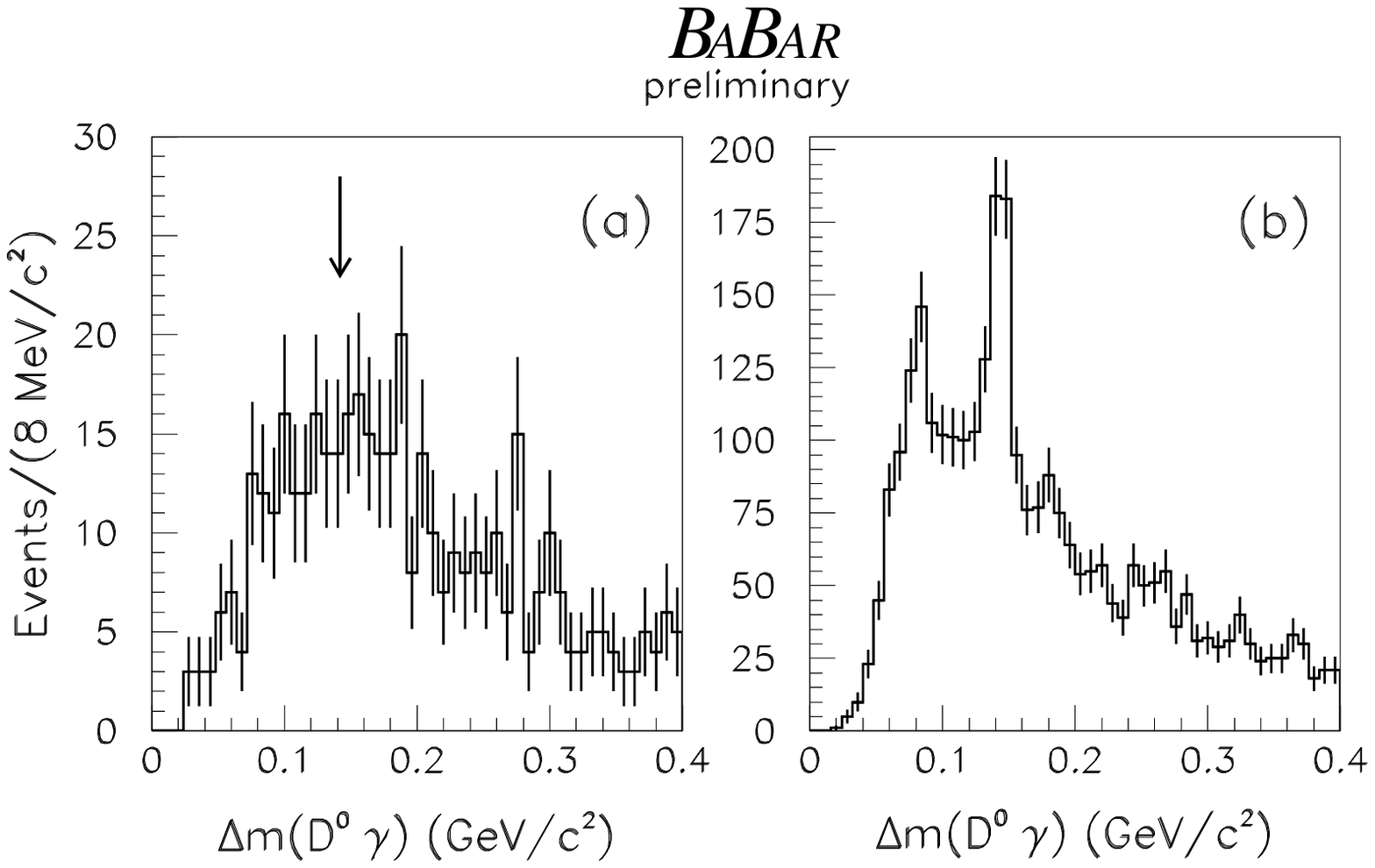}
\caption{Spectra of $\Delta m(\Dz \gamma) \equiv m(\Dz \gamma) - m(\Dz) $ for (a) all events inside and
(b) $\Dz \to K^- \pi^+, \Dzb \to K^+ \pi^-$  events outside the exclusive ISR region. 
The arrow indicates the expected position of $D^{*0}$.}
\label{fig:fig3}
\end{center}
\end{figure}

\begin{table}[tbp]
\caption{Mass resolutions and event yields for each reconstructed channel.}
\label{tab:tab1}
\begin{center}
\vskip -0.2cm
\begin{tabular}{lccccc}
\hline
 & & & & & \cr
Channel & $D$ Mass Res. & Candidates & Background & Signal & $D \bar D$ Mass Res.\cr
 & (\mevcc) & & & & (\mevcc)\cr
\hline
$\Dz \to K^- \pi^+, \Dzb \to K^+ \pi^-$ & 6.8 $\pm$ 0.2 & 68 & 4 & 64 $\pm$ 8 & 5.1 $\pm$ 0.1 \cr
$\Dz \to K^- \pi^+ \pi^0, \Dzb \to K^+ \pi^-$ & 13.0 $\pm$ 0.4 & 121 & 22 & 99 $\pm$ 12 & 5.2 $\pm$ 0.2 \cr
$\Dz \to K^- \pi^+ \pi^+ \pi^-, \Dzb \to K^+ \pi^-$ & 5.0 $\pm$ 0.04 & 111 & 16 & 95 $\pm$ 11 &
4.6 $\pm$ 0.1\cr
$D^+ \to K^- \pi^+ \pi^+, \Dzb \to K^+ \pi^- \pi^- $ & 5.6 $\pm$ 0.1 & 116 & 14 & 102 $\pm$ 11 & 4.6 $\pm$ 0.1 \cr
\hline
\end{tabular}
\end{center}
\end{table}

A potential source of contamination in this sample of $e^+e^- \to  \gamma_{ISR} D \bar D$
candidates could arise 
from the low $MM^2$ tail from $e^+e^- \to  \gamma_{ISR} D^{(*)} \bar D^{(*)}$. This background is partially 
suppressed by the extra-$\pi^0$ veto; however radiative $D^{*0}$ decays or $D^{*0} \to \pi^0 D^0$ decays 
with an unreconstructed photon daughter of the $\pi^0$ could survive. A direct search for $D^{*0}$ 
is made by plotting  $\Delta m(\Dz \gamma) \equiv m(\Dz \gamma) - m(\Dz)$ 
in Fig.~\ref{fig:fig3}(a) for all $\Dz$ and $\Dzb$ candidates within the ISR $MM^2$ region.
For comparison, the same quantity is shown for 
$\Dz \to K^- \pi^+, \Dzb \to K^+ \pi^-$ candidates outside the 
ISR $MM^2$ region in Fig.~\ref{fig:fig3}(b). A clear peak in Fig.~\ref{fig:fig3}(b) 
corresponding to $D^{*0} \to \gamma D^0$ is visible,
but there is no evidence for any $D^{*0}$ contamination in the ISR $D \bar D$
sample.

\section{$D \bar D$ mass spectrum} 
\begin{figure}[!htb]
\begin{center}
\includegraphics[width=12cm]{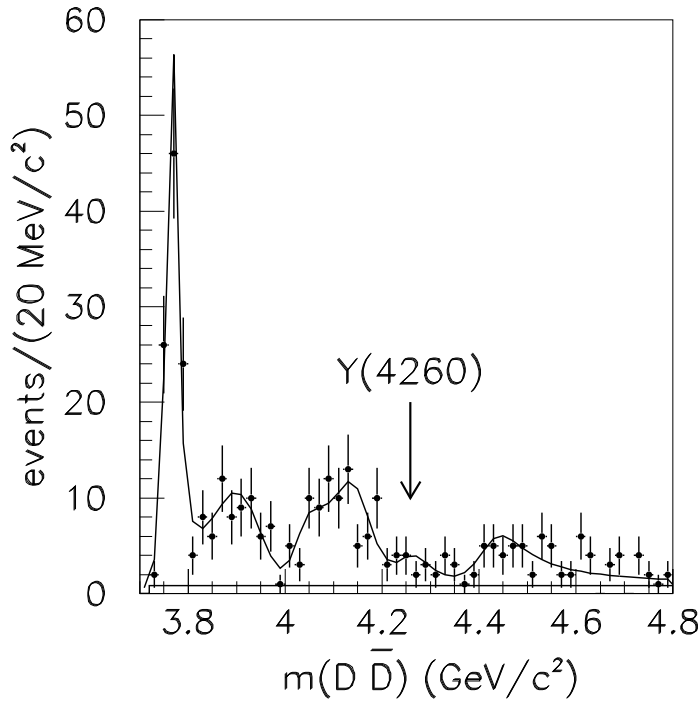}
\caption{The $D \bar D$ invariant mass spectrum, summed over all four reconstructed final states,
with a fit that includes the $Y(4260)$ contribution. 
The arrow indicates the expected position of the $Y(4260)$.}
\label{fig:fig4}
\end{center}
\end{figure}

The $D \bar D$ invariant mass spectrum for the ISR sample is shown by the data points in 
Fig.~\ref{fig:fig4}. A clear signal is seen for the $\psi(3770)$, which decays 
predominantly to $D \bar D$. Additional enhancements coincide with the $\psi(4040)$, $\psi(4160)$ 
and $\psi(4415)$ masses, which are observed in measurements of the total $e^+e^-$ hadronic 
cross section. Finally, a broad enhancement is evident near 3.9 \gevcc. This structure 
has not been seen in the hadronic cross section measurements but is in qualitative agreement with 
coupled-channel model predictions~\cite{eichten}. 

An unbinned maximum likelihood fit is performed 
using a signal shape described by four relativistic P-wave Breit-Wigner distributions 
convoluted with a P-wave phase space function. The Breit-Wigner parameters are fixed to 
the values from a fit to the hadronic cross section~\cite{seth}. A Gaussian term
is used to parameterize the enhancement near 3.9 \gevcc.
All Breit-Wigner and Gaussian terms are allowed to interfere by assigning them a free 
phase. A constant background is 
fixed to a value of $0.84$ events per 20 $\mevcc$ bin 
as determined  from a fit to the $D$ and $\bar D$ mass sidebands. 
The fit yields a Gaussian-term mass of 
($3.909 \pm 0.021$) \gevcc with $\sigma = (0.050 \pm 0.007$) \gevcc.
The $D \bar D$ mass resolution at the $Y(4260)$ mass, 
determined from Monte Carlo studies and shown in Table~\ref{tab:tab1}, is small 
compared to the widths of the fit structures and is neglected. 

We consider a possible $Y(4260)$ contribution 
by performing a second fit, adding
a fifth Breit-Wigner term with parameters determined from the $Y(4260)$ 
observation in the $J/\psi \pi^+ \pi^-$ mode: 
$m(Y(4260)) = 4.260$ \gevcc and $\Gamma(Y(4260)) = 0.088$ \gevcc~\cite{y4260}. The 
Gaussian term is fixed to parameters from the first fit. The result, shown in
Fig.~\ref{fig:fig4}, yields 
a $Y(4260)$ component with $7 \pm 13$ events. The 
systematic uncertainty is evaluated as $\pm 8$ events by varying the mass and width of the four
$\psi$ 
resonances, the $Y(4260)$, and the Gaussian enhancement near 3.9 \gevcc by one standard deviation, 
and by removing the phase space factor from the fit. All the resulting deviations are added in quadrature. 

\begin{figure}[!htb]
\begin{center}
\includegraphics[width=10cm]{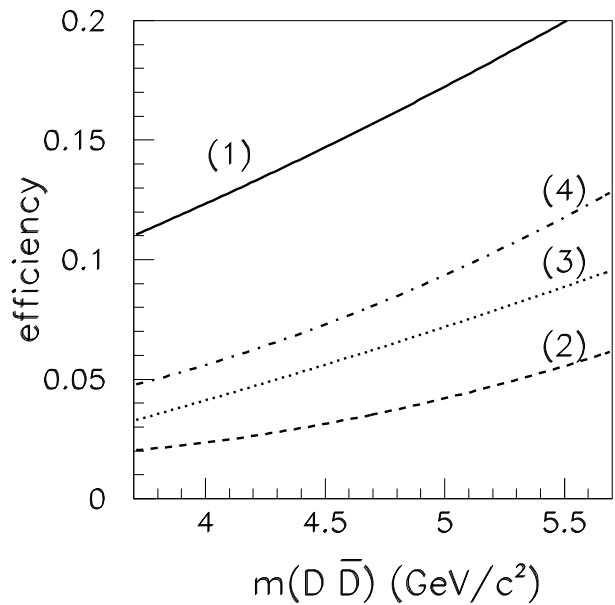}
\caption{Reconstruction efficiency for the four channels as defined in Eqs.\ (1)-(4) 
as a function of the $D \bar D$  invariant mass.}
\label{fig:fig5}
\end{center}
\end{figure}

The $D \bar D$ reconstruction efficiency for each channel, determined from Monte Carlo 
studies, increases with $D \bar D$ mass as shown in Fig.~\ref{fig:fig5}. 

Combining the fit result with world average
$D$ branching fractions, reconstruction efficiency, and the result from the 
\babar~$\gamma_{ISR} \psi \pi^+\pi^-$ analysis~\cite{y4260} yields 
\begin{equation}
\frac{\BR(Y(4260)\to D \bar D)}{\BR(Y(4260)\to J/\psi \pi^+ \pi^-)} = 1.4 \pm 3.1,
\end{equation}
where statistical and systematic uncertainties have been added in quadrature. Because a 
statistically significant signal for $Y(4260)\to D \bar D$ has not been established, this result
 can be recast as an upper limit:
\begin{equation}
\frac{\BR(Y(4260)\to D \bar D)}{\BR(Y(4260)\to J/\psi \pi^+ \pi^-)} < 7.6~\mathrm{at~ 95\%~C.L.}
\end{equation}
This limit is over an order of magnitude smaller than the value found for the $\psi(3770)$, 
another indication that the $Y(4260)$ is not a conventional vector charmonium state. 
\section{Conclusions}
In summary, ISR events have been used to
explore $D \bar D$
production in $e^+e^-$ annhilation from charm threshold to 6 \gevcc. We find evidence for 
$D \bar D$ decays of the $\psi(3770)$, $\psi(4040)$, $\psi(4160)$, and 
$\psi(4415)$, 
and a broad enhancement of $D \bar D$ production near 3.9 \gevcc in qualitative agreement with coupled-channel model predictions. No statistically significant signal for $Y(4260) \to D \bar D$ is seen, 
leading to a 95\% confidence level upper limit, $\frac{\BR(Y(4260)\to D \bar D)}{\BR(Y(4260)\to J/\psi \pi^+ \pi^-)} < 7.6$.

\section{Acknowledgments}
\label{sec:Acknowledgments}

We are grateful for the 
extraordinary contributions of our \pep2\ colleagues in
achieving the excellent luminosity and machine conditions
that have made this work possible.
The success of this project also relies critically on the 
expertise and dedication of the computing organizations that 
support \babar.
The collaborating institutions wish to thank 
SLAC for its support and the kind hospitality extended to them. 
This work is supported by the
US Department of Energy
and National Science Foundation, the
Natural Sciences and Engineering Research Council (Canada),
Institute of High Energy Physics (China), the
Commissariat \`a l'Energie Atomique and
Institut National de Physique Nucl\'eaire et de Physique des Particules
(France), the
Bundesministerium f\"ur Bildung und Forschung and
Deutsche Forschungsgemeinschaft
(Germany), the
Istituto Nazionale di Fisica Nucleare (Italy),
the Foundation for Fundamental Research on Matter (The Netherlands),
the Research Council of Norway, the
Ministry of Science and Technology of the Russian Federation, and the
Particle Physics and Astronomy Research Council (United Kingdom). 
Individuals have received support from 
the Marie-Curie IEF program (European Union) and
the A. P. Sloan Foundation.

\end{document}